\documentclass[letterpaper]{aa}
\usepackage{graphicx}
\usepackage{txfonts}

\begin{document}

\def\te{T_{e\hskip-0.5pt f\hskip-1.0pt f}}
\def\delt{\scriptstyle \Delta }

\def\eol{\hfil\break}
\def\va{\vskip 5.0 mm} \def\vb{\vskip 2.5 mm} \def\vc{\vskip 1.5 mm}
\def\ha{\hskip 5.0 mm} \def\hb{\hskip 2.5 mm} \def\hc{\hskip 1.5 mm}
\def\vd{\vskip 1.0 mm} \def\hd{\hskip 1.0 mm}

   \title{Catalogue of averaged stellar effective magnetic fields. }
   \subtitle{I. Chemically peculiar A and B type stars}
   \author{V.D. Bychkov\inst{1,2}, L.V. Bychkova\inst{1},
	   J. Madej\inst{3} }

   \offprints{V.D. Bychkov,   \email{vbych@sao.ru }}

   \institute { $^1$ Special Astrophysical Observatory of the Russian
       Academy of Sciences, Nizhnij Arkhyz, 369167 Russia \\
    $^2$ Isaac Newton Institute of Chile, SAO Branch, Nizhnij Arkhyz,
	 369167 Russia \\
    $^3$ Astronomical Observatory, University of Warsaw,
    Al. Ujazdowskie 4, 00-478 Warszawa, Poland  }

   \date{Received ...; accepted ...}

   \abstract{

This paper presents the catalogue and the method of determination of
averaged quadratic effective magnetic fields $\langle B_e \rangle$ for
596 main sequence and giant stars. The catalogue is based on measurements
of the stellar effective (or mean longitudinal) magnetic field strengths $B_e$, which were
compiled from the existing literature.

We analysed the properties of 352 chemically peculiar A and B stars in the catalogue,
including Am, ApSi, He-weak, He-rich, HgMn, ApSrCrEu, and all ApSr type stars.
We have found that the number distribution of all chemically peculiar (CP) stars {\em vs.}
averaged magnetic field strength is described by a decreasing exponential
function. Relations of this type hold also for stars of all the analysed
subclasses of chemical peculiarity. The exponential form of the above distribution
function can break down below about 100 G, the latter value representing
approximately the resolution of our analysis for A type stars.


}
   \maketitle

%

\section{Introduction}

\rm
Research on stellar magnetic fields is among the most important
issues in both observational and theoretical astrophysics.
The first measurements of magnetic fields in stars were done more
than 50 years ago (Babcock \& Burd 1952). From that time, both the
number of magnetic field measurements and the number of investigated
stars have grown enormously. Therefore we decided to collect and present
in some homogeneous form all the published magnetic field measurements.
We have also attempted to analyse these preprocessed observational data.

Similar efforts have been made previously, but were based on much less numerous
sets of measurements (Brown et al. 1981; Borra et al. 1983; Glagolevskij
et al. 1986; Bychkov 1990; Bychkov et al. 1990). The above compilations
have been essential for our understanding of the magnetic field strength and
structure in stellar atmospheres, and their generation and time
evolution in stellar interiors. Taking into account the large increase of
the accumulated observational material, we believe that analogous new
research of this kind is necessary and fully justified.

The catalog and analyses presented below do not include either
isolated degenerate stars (cooling neutron stars and most white dwarfs),
or degenerate stars in interacting binaries. Only a few of the brightest white
dwarfs are present in the catalog.

\section{Averaging of stellar effective magnetic fields }

The differential contribution $dB_e$ to the effective magnetic field of a star
is defined as the area-weighted projection of the local vector of the magnetic field
${\bf B}_{\, loc}$ onto the line of sight. The local monochromatic intensity
$I_\nu$ of outgoing radiation is also a weighting factor in that projection.
The effective (or mean longitudinal) magnetic field $B_e$ is the weighted mean value, integrated over
the visible stellar disc
\begin{equation}
 B_e = { \int\limits_0^{2\pi} \int\limits_0^{\pi/2} B_{\, loc} \cos \gamma \,
I_\nu (\theta) \sin \theta \cos \theta \, d\theta \, d\varphi\over
{\int\limits_0^{2\pi} \int\limits_0^{\pi/2}  \,
I_\nu (\theta) \sin \theta \cos \theta \, d\theta \, d\varphi}} \> ,
\end{equation}
where $\gamma$ denotes the angle between the local vector ${\bf B}_{\, loc}$
and the direction towards the observer. The variable $\theta$ denotes the
colatitude angle, and $\varphi$ stands for the azimuthal angle of the angular
integration. The above definition assumes a simplified situation, in which
the $B_e$ is determined at a single discrete frequency only (Madej 1983).

In general, the specific intensity of radiation
$I_\nu (\theta)$  depends strongly on the frequency of radiation $\nu$, and
exhibits various limb-darkening relations for different $\nu$s. Therefore
the value of the effective magnetic field $B_e$ is also a frequency dependent
quantity, when measured for the given magnetic field configuration of
a star.

The dependence of $B_e$ on frequency, or on the finite range of frequencies
in which measurements were done, has  always been ignored in earlier papers,
which are collected and analysed here. Therefore also in this paper we
do not distinguish $B_e$ values measured in the wings of the hydrogen
Balmer lines, or elsewhere in the optical spectra of stars.

In most magnetic stars the values of $B_e$ change periodically with the
rotational phase of the star. Values of $B_e$ can be either positive or
negative. Moreover, it is possible that a star with strong magnetic field
can momentarily exhibit $B_e = 0$, depending on the aspect.
Therefore it is useful to characterize the magnetic properties of various
stars by the averaged quadratic effective magnetic field
$\langle B_e \rangle$, which is always positive (Borra et al. 1983).

For a series of  $B_{e}$ measurements, we define
\begin{eqnarray}
\langle B_e \rangle &=& \left( {1\over n} \sum\limits_{i=1}^n B_{ei}^2
   \right)^{1/2}\, ,\\
\langle \sigma_e \rangle &=& \left( {1\over n} \sum\limits_{i=1}^n
   \sigma_{ei}^2  \right) ^{1/2} \, ,
\label{eq:fer1}
\end{eqnarray}
where $B_{ei}$ denotes the $i$-th measurement of the effective
magnetic field, and $n$ is the total number of observations for a given
star. The variable $\sigma_{ei}$ is the standard error of $B_{ei}$, and
$\langle \sigma_e \rangle $ is the rms standard error of $\langle B_e
\rangle $.

The value of $\chi^2/n$ (given per single degree of freedom) allows one to
judge whether a series of $B_{ei}$ for a given star represents a reliable detection of a nonzero
effective magnetic field, or or whether this series is rather the result of random noise
\begin{equation}
  \chi^2/n = {1\over n} \sum\limits_{i=1}^n  {B_{ei}^2 \over
    \sigma_{ei}^2 }   \, .
\end{equation}

This method for averaging the individual $B_e$ measurements of
a magnetic star was introduced by Borra et al. (1983), to study magnetic
properties of He-weak stars. This evaluation of $\langle B_e \rangle $
is particularly useful to study stars with few or high
noise $B_e$ observations, where full magnetic curves cannot yet be constructed.

Borra et al. (1983) have pointed out that the value of $\langle B_e \rangle$
gives an estimate of the amplitude of the $B_e$ variations of a given star,
provided that this amplitude is substantially larger than $\langle \sigma_e \rangle $.

\section{Description of the tables}

Descriptions of stars and the available magnetic field data about each are included in a series of
10 tables. The basic and most extensive Table A.1, included in Appendix A,
presents the full listing of stars for which we performed computations of the
quadratic $\langle B_e \rangle $ averages. For convenience, these
stars are ordered according to their HD number. Successive rows of Table A.1 give:
HD number (or BD number in case of faint stars), spectral type, number $N$ of magnetic observations, value of
$\langle B_e \rangle $ in G, standard deviation $\sigma$ in G, value of
$\chi^2/n$, method of $B_e$ determination (abbreviations are explained at
the bottom of Table A.1), and numbers referring to papers where we found
the original magnetic field measurements. Cross-references between these
numbers and the original papers are also given at the bottom of Table A.1.

Table A.1 contains magnetic data on a total of 596 stars of various
spectral types. One can easily see that in the case of many stars listed
there, the value of $\langle B_e \rangle$ is approximately equal or smaller
than $\langle \sigma_e \rangle$, which usually means that detection of the magnetic
field itself highly uncertain.

Table 1 summarizes our results on the distribution of averaged effective
magnetic fields in Ap type stars of various subclasses. In this table, the number $N$ in the second column displays
the number of stars of a given peculiarity type in our sample.

One should note that some CP stars exhibit more than one type of chemical
peculiarity simultaneously. For example, it is a well known observational
fact that some Si-type stars appears also as He-weak stars, etc. When this happened, we have included such ambiguous stars in both samples.
Consequently, the sum of all Ap stars (352) is lower than the number of
stars summed over all particular types of peculiarity in Table 1.

Tables 2--9, which are put in the main body of the paper, present lists
of the 352 Ap stars of the sample distributed into various types of spectral peculiarity.
The tables present individual stars and give for each of them: HD number,
HR number, name of the star (or BD number), and spectral type including
the type of peculiarity. No magnetic field data are listed here.

Due to enormous complexity of Sr-like chemical peculiarities which have been
identified in some Ap stars, we have attempted to separate 136 stars
which exhibit essentially only the SrCrEu spectral type. These stars are
listed in Table 8. Complementing this table, Table 9 presents list of the other 43 Sr-type
stars which exhibit SrCrEu type mixed with other peculiarities.
The logical sum of Tables 8 and 9 forms the class ``Sr all'', which contains
179 stars and is a single entry in Table 1.

The stars listed in Tables 2--9 are exactly the objects which were used
to construct Figures 1--9, and to obtain number distribution functions
of the averaged quadratic magnetic fields $\langle B_e \rangle$ for each
types of peculiarity considered among the A type stars on the main sequence.

\subsection{Reevaluation of $\langle B_e \rangle $ errors }

For many years observers have always estimated the {\em standard error} of
effective magnetic field measurements. However,
some early papers tabulated probable errors of $B_e$, which
should be transformed to standard errors to ensure their compatibility.

Moreover, some of the early $\langle B_e \rangle $ measurements have
unrealistically small standard errors of the order of a few tens of G.
This comment refers mostly to photographic magnetic field observations;
see papers by Babcock et al. in the list of references, for example.

An independent error estimate of published $B_e$ determinations
can be obtained by one of the following methods.

\noindent
{\bf 1.} Consider a star with no apparent $B_e$ variations. In the case where
we have a sufficiently large set of $B_e$ measurements, we can simply
determine the mean $\langle B_e \rangle$ value and the error of a single
$B_e$ measurement in the standard manner.

\noindent
{\bf 2.} If the magnetic field $B_e$ varies with time, and the parameters
of phase variability are unknown, then we can estimate the
$\langle B_e \rangle$ and the upper limit of error of a single measurement
in the same standard way. In this case the observed scatter of individual
$B_e$ observations include both real errors plus the unknown magnetic field
variability.
The lower the contribution of $B_e$ variability is to the scatter, the more
realistic the error estimates are.

\noindent
{\bf 3.} If the magnetic field varies with time, and if the parameters of
(periodic) phase variability are known, then we simply determine the
mean $B_e (\varphi)$ phase curve and compute the predicted magnetic
field strength corresponding to each observed point. Finally, we determine
the error of single measurement as was done in paragraph {\bf 1}.

\vskip 2 mm

The general considerations presented above should be supplemented by the
following comments:

\noindent
{\bf 4.} The averaged value of the effective field, $\langle B_e \rangle$,
significantly depends on the choice of useful spectral lines.
This is particularly important for early observations, since then
analysing instruments worked in narrow spectral windows (200 - 300 {\AA }),
and the number of lines used for the $B_e$ determination was very small.

\noindent
{\bf 5.} The best average $B_e (\varphi)$ curves were obtained analysing
Zeeman splitting of the Balmer lines. However, $B_e$ measurements obtained
from Balmer lines and metallic lines can differ substantially due
to the well-known effect of inhomogeneous distribution of elements
over the surface of a magnetic star.

\noindent
{\bf 6.} The accuracy of $B_e$ measurements depends not only on the particular
set of spectral lines and their total number, but also on the apparent
rotational broadening, i.e. on ${\rm v} \sin i$. Rotational broadening of
lines strongly influences the accuracy.

\vskip 2 mm

In order to obtain reliable error estimates of older $B_e$
measurements, we have selected 21 stars observed by the following
authors: H.W. Babcock, G.W. Preston, S.C. Wolff. and W.K. Bonsack.
Those stars were not necessarily observed by all four of them.
We also took into account papers in which they were present in the
author's list (cf. Bibliography in this paper).

The set of 21 stars consists of (HD numbers): 9996, 18296, 24712, 32633,
62140, 65339, 71866, 74521, 112413, 118022, 125248, 133029, 137909,
152107, 153882, 168733, 175362, 187474, 188041, 196502, 201601, and
215441.
For all of them projected rotational velocities ${\rm v} \sin i$ are known.
Numerous $B_e$ determinations for these stars were obtained both with
the ``old'' photographic technique, and with new high-accuracy methods.
Details of our error calibrations of the earliest $B_e$, and its
dependence on both ${\rm v} \sin i$ and the effective temperature $\te$,
will be presented in a forthcoming paper.

From our analysis, we obtained the following standard errors of $B_e$,
corresponding to photographic observations of the most active authors:

\halign { \ha #\ha \hfil & # \cr
  Babcock, H.W. & $\sigma = 237.6\times \exp(0.0465 \> {\rm v}\sin i) $ \cr
  Wolff, S.C.   & $\sigma = 196.4\times \exp(0.0369 \> {\rm v}\sin i) $ \cr
  Preston, G.W. & $\sigma = 194.9\times \exp(0.0558 \> {\rm v}\sin i) $ \cr
  Bonsack, W.K. & $\sigma = 151.3\times \exp(0.0610 \> {\rm v}\sin i) $ \cr
}

\noindent
The standard error $\sigma$ is given in G, and velocity ${\rm v} \sin i$ is
in km/sec. As is evident, the accuracy of $B_e$ measurements increases from
the earliest observations by H.W. Babcock to later data by W.K. Bonsack.

We have recomputed errors of $B_e$ measurements of the above observers with
the above relations in all cases in which their original errors were not
given or were apparently unrealistically small. In this way have ensured
compatibility of the earliest and modern $B_e$ determinations.

\section{Distribution of averaged effective magnetic fields}

Let us substitute $B = \langle B_e \rangle$ for brevity. From the
data collected in Table A.1 and in Tables 2--9 we
have constructed two types of relations. They display the dependence of
the number distribution function $N(B)$, and its integral over $B$, on
the average effective magnetic field $B$ of the A type stars.

Quantity $N(B) \, dB$ gives the number of stars in a given group having
the average quadratic effective magnetic field $B$ in the range
$\left[ B, B+dB \right]$.

\subsection{Integrated distribution function}

We define the integrated distribution function as
\begin{equation}
 N_{Int} (B) = N_{tot} - \int \limits_0^B  N(B') \, dB' \, ,
\end{equation}
where $N_{tot}$ denotes the total number of stars belonging to that group.

We have investigated separately $N(B)$ for Am, He-weak, He-rich, Si, HgMn,
SrCrEu, all Sr-type, and all stars displaying Hg or Mn. Discussion of the
distribution function for stars of other spectral types has been deferred
to the following papers.

For a given subclass of stars, we have divided the range of the quadratic
averaged magnetic field $\langle B_e \rangle$ from zero to the maximum
field in this group into up to 40 bins of equal length (80 bins for Si stars),
and counted the number of stars in each bin. Figures 1--9 display the
relation between the discrete $\langle B_e \rangle$
and the summed number of stars located in higher bins, expressed in
percent of the total number of stars in that peculiarity class. Such a
relation represents the integrated distribution function $N_{Int}(B)$,
and describes the probability that upon investigating a new star of this
chemical peculiarity, its $\langle B_e \rangle$ will be higher than the
value of $B$. That relation is given by series of dots in the Figures.

Figures 1--9 demonstrate the striking rule, that all the corresponding
functions $N_{Int} (B)$  are well approximated by the exponential function,
normalized to unity at $B=0$
\begin{equation}
  N_{Int} (B) / N_{tot} = {a_1 \over {100 \%}} \, \exp \, (-B/a_2) \, .
\end{equation}
Coefficients $a_1$ (in \%) and $a_2$ (in G) depend on the class of chemical
peculiarity.

Columns 3-4 of Table 1 present the best fit coefficients $a_1$ and $a_2$
for all the analysed subclasses. The last three columns present the fixed
values of magnetic field intensity $\langle B_e \rangle^{fix}$ (in G),
defined that the number of stars with
$\langle B_e \rangle \ge \langle B_e \rangle^{fix}$ constitutes 30\%,
50\%, and 70\% of the total number of stars belonging to this peculiarity
type, cf. column 2. In fact, these three columns are just some exemplary
solutions of Eq. 6 for the magnetic field $B$.

\begin{table}
\caption{ Best fit parameters }
\label{tab: 1}
\renewcommand{\arraystretch}{1.1}
\begin{tabular}{|l|r|r|r|r|r|r|}
    \hline
Peculiarity& N	&$a_1 (\%) $&$a_2$ (G)& 30\%& 50\%& 70\% \\
\hline
all Ap   &  352&   97.2&  789.2&  928&  525&  259  \\
Sr all   &  167&  106.9& 1081.2& 1360&  819&  448  \\
Sr only  &  126&  108.6& 1018.1& 1310&  790&  447  \\
Am       &   44&   95.3&  110.5&  127&   71&   34  \\
He-rich  &   19&   97.2&  916.1& 1080&  609&  301  \\
He-weak  &   60&  116.0&  717.9&  970&  604&  363  \\
Hg \& Mn &   39&   74.9&  515.7&  471&  208&   34  \\
HgMn only&   19&   75.2&  350.1&  322&  143&   25  \\
Si       &  159&  102.0&  906.1& 1110&  646&  341  \\
\hline
\end{tabular}
\begin{list}{}{}
\item[$^{\mathrm{a}}$] Some of the stars exhibit few different
peculiarity types simultaneously, and they are counted in more
than one row of Table 1.
\end{list}
\end{table}

\subsection{The distribution function}

The distribution function can immediately be obtained from $N_{Int}$ by
the relation
\begin{equation}
  N(B) = - {dN_{Int}\over {dB}}  \, .
\end{equation}
The function $N(B)$ is therefore also an exponential function with the above
analytic approximation
\begin{equation}
  N(B) =  N_{tot} \> {a_1 \over{100 \%}} \> a_2^{-1} \,
          \exp \, (-B/a_2) \, .
\end{equation}
If one attempts to construct the distribution function in direct way,
based on the tabulated data, then this function would exhibit serious
noise due to the limited number of data points.

One should note that the above exponential dependence has been determined
using a  resolution ${\delt} B \approx 100$ G only, which corresponds
to average size of single bin in Figs. 1--9. This implies that with our
method of figure construction, we cannot say anything about the shape of the
distribution function in the region of the weakest magnetic fields $B$
below $\approx 100$ G. Note that the value of ${\delt } B$ resolution mentioned above is
averaged over all spectral subclasses, in fact, it is in
the range ${\delt } B = 25-200$ G (see the Figures).

It is important to stress here, that we have set the total number of stars
of a given subtype $N_{tot}$ to 100\%, no matter whether the given stars had
detectable magnetic fields or not.

   \begin{figure}
    \resizebox{\hsize}{!}{\rotatebox{0}{\includegraphics{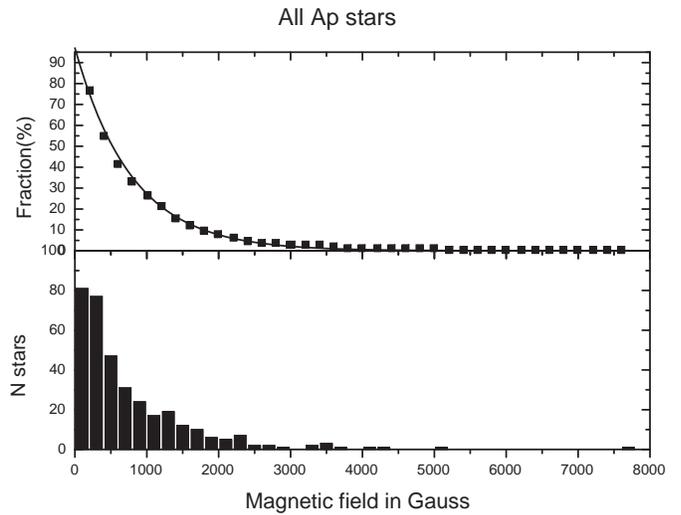}}}
    \caption[]{Integrated distribution function $N_{Int}(B)$ in percent
      (upper panel), and the number distribution function $N(B)$ (lower
      panel) for all Ap stars. }
    \label{fig:fig1}
   \end{figure}

   \begin{figure}
    \resizebox{\hsize}{!}{\rotatebox{0}{\includegraphics{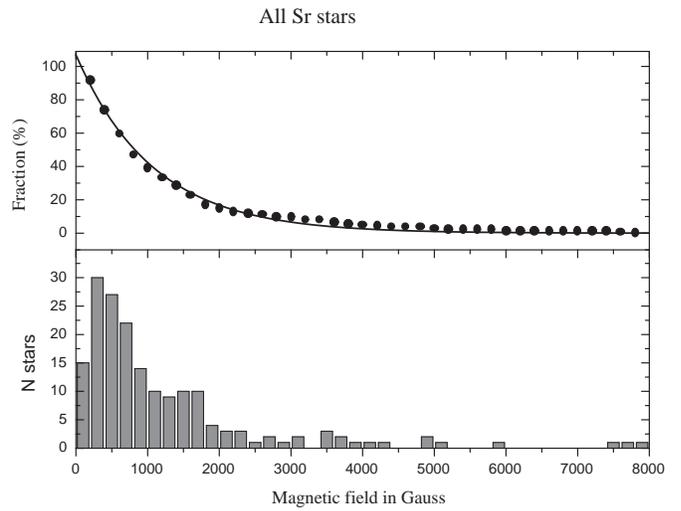}}}
    \caption[]{ The same for all Sr type stars, also those with mixed
       peculiarities.  }
    \label{fig:fig1}
   \end{figure}

   \begin{figure}
    \resizebox{\hsize}{!}{\rotatebox{0}{\includegraphics{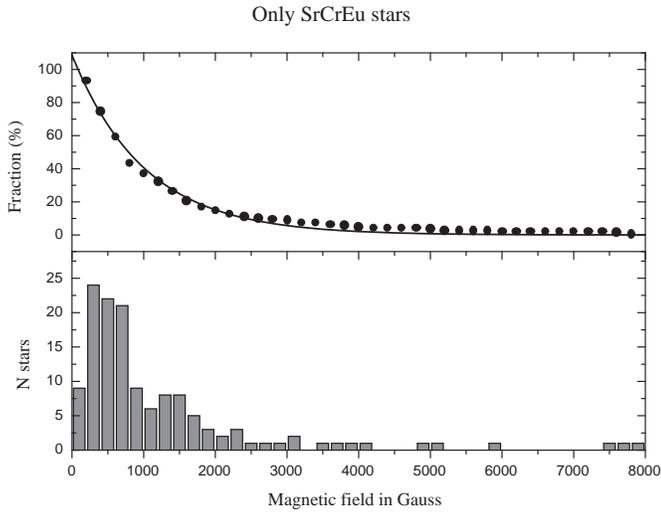}}}
    \caption[]{The same for SrCrEu stars.  }
    \label{fig:fig2}
   \end{figure}

   \begin{figure}
    \resizebox{\hsize}{!}{\rotatebox{0}{\includegraphics{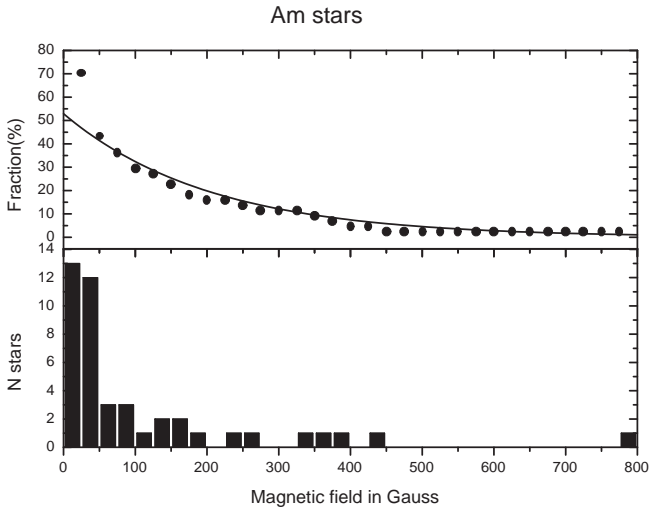}}}
    \caption[]{The same for Am stars.  }
    \label{fig:fig3}
   \end{figure}

   \begin{figure}
    \resizebox{\hsize}{!}{\includegraphics{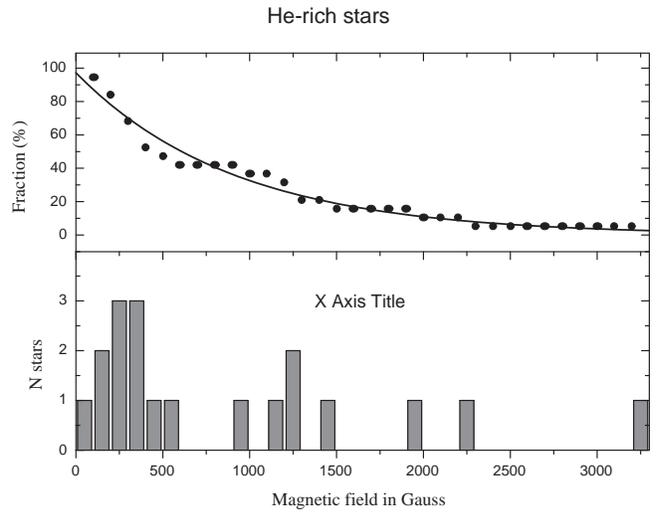}}
    \caption[]{The same for He-rich stars.  }
    \label{fig:fig4}
   \end{figure}

   \begin{figure}
    \resizebox{\hsize}{!}{\includegraphics{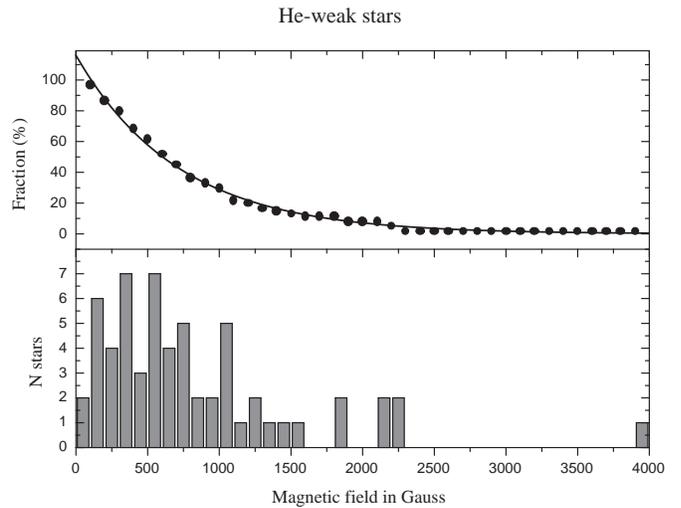}}
    \caption[]{The same for He-weak stars.  }
    \label{fig:fig5}
   \end{figure}

   \begin{figure}
    \resizebox{\hsize}{!}{\includegraphics{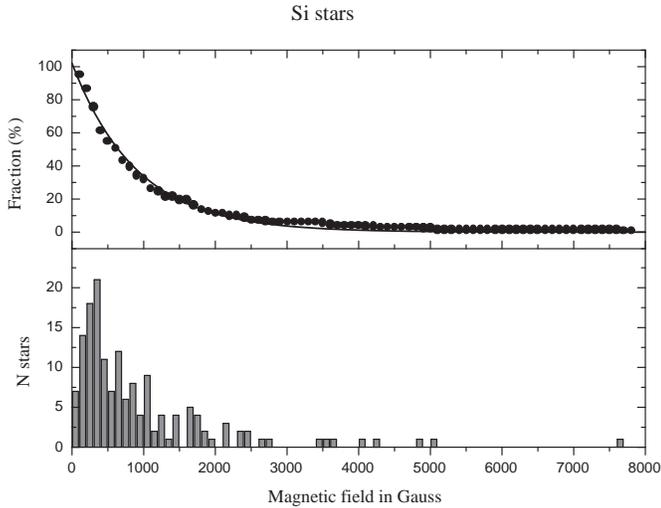}}
    \caption[]{The same for Si stars.  }
    \label{fig:fig6}
   \end{figure}

   \begin{figure}
    \resizebox{\hsize}{!}{\includegraphics{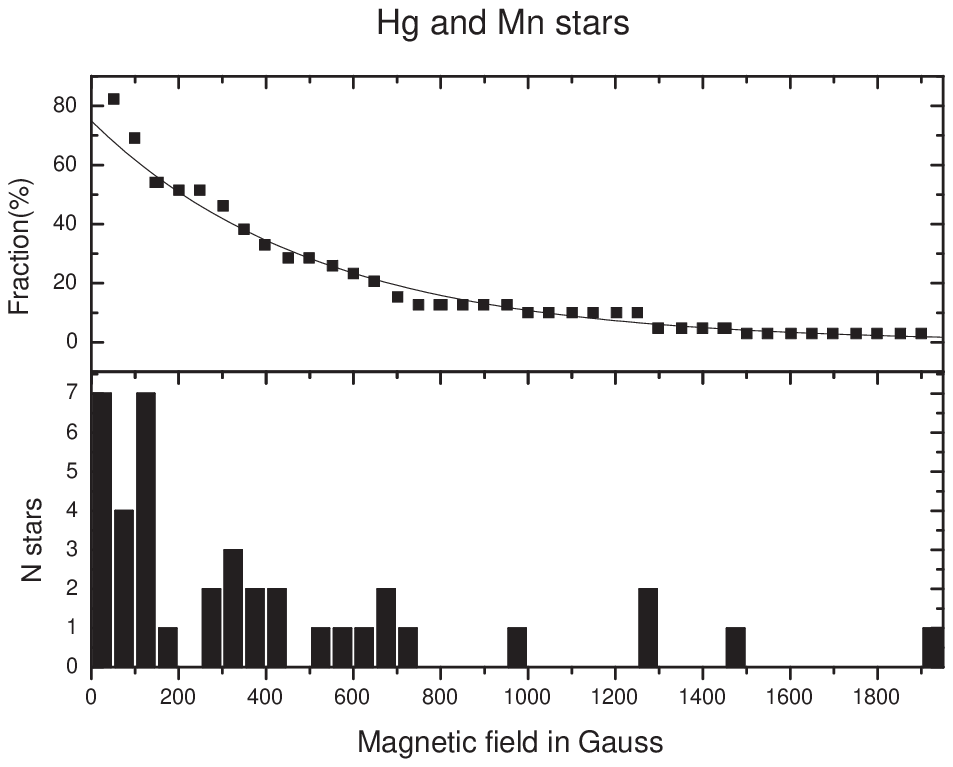}}
    \caption[]{The same for all stars including Hg or Mn, also of mixed peculiarities.  }
    \label{fig:fig7}
   \end{figure}

\subsection{Distortion of the distribution functions }

The referee pointed out that our $\langle B_e \rangle$ and $\langle
\sigma_e \rangle$ statistics, and the distribution functions $N_{int}$
presented in this paper, can be distorted due to the following reasons.

{\bf 1.}
Errors of $B_e$ measurements taken by different observers and techniques
sometimes strongly differ. In such case average values of $\langle B_e
\rangle$ and $\langle \sigma_e \rangle $ in Table A.1 can be inflated by
few very inaccurate measurements with large individual $\sigma_e$,
cf. Eqs 2--3. This is particularly important for stars with weak magnetic
fields, for which the number of available $B_e$ observations is small
(e.g. the Am star 68 Tau).

The most accurate $B_e$ measurements should weight mostly when computing
averages $\langle B_e \rangle$ and $\langle \sigma_e \rangle$. However,
Eqs 2--3 defined by Borra et al. (1983) assign the same weight of unity
to all $B_e$ measurements; i.e. their $\langle B_e \rangle$ and $\langle
\sigma_e \rangle$ statistics are most meaningful when they resulted from
data with comparable errors. In the present paper we follow strictly
the above definitions, Eqs 2--3, and did not alter them in any way e.g.
by eliminating $B_e$ data of outstanding $\sigma_e$ errors.

{\bf 2.}
The above effect implies also, that the distributions of $N_{int}$ are
certainly distorted by inclusion of stars for which $\langle B_e \rangle$
has been exaggerated. This is particularly important at low field end of
$N_{int}$ for all subclasses of Ap stars, and for the whole group of
essentially nonmagnetic Am stars, see Fig. 4 (see also the following
subsection). Therefore runs of $N_{int}$ derived in this paper certainly
are distorted below $\approx$ 100 to 300 G, or so,

In spite of this effect we believe that even for Am stars constructing our
$N_{int}$ makes sense, because it represents the upper limit constraining
their true distribution, $N_{int}^{true}$.

\subsection{Low magnetic fields}

One can easily see that the distribution function $N(B)$ exhibits
a significant drop of the averaged quadratic field at the limit
$B \rightarrow 0$. Such behaviour can be seen for almost all of investigated
classes of chemical peculiarities, with the exception of Am stars only.

The origin of this behavior of the directly measured $N(B)$ cannot be
explained with full confidence. On one hand, the numbers of star counts in
$B$ bins in Figures 1--9 is very low, and therefore strong random
fluctuations are very likely. On the other hand, we believe that such
asymptotic drops of $N(B)$ are rather due to random errors of the directly
measured effective magnetic fields $B_e$. The quadratic average of errors
${\delt } B_e$ is not likely to approach 0 G, particularly for poorer
observations, and it is comparable with the width of a bin. Therefore a low
number of stars with quadratic field $B\approx 0$ represents just some
type of statistical selection effect.

As was pointed out by the referee, the observational data analysed here
for magnetic star calsses probably exhibit a deficiency of stars with low magnetic fields.
This is because observers frequently were not interested in observing
stars in which the intrinsic magnetic field appeared to be small,
and stars with stronger fields were always favoured.
Such a personal bias certainly distorts the observed distributions $N_{Int}$
in each subclass of A type stars, which are convolutions of intrinsic
distributions with an ``observer interest'' distribution. The effect is
very difficult, if not impossible, to  correct in general. We
believe, however, that the effect influences counts $N_{Int}$ only in
the lowest bins of our histograms, which are underpopulated also due to
reasons discussed in the previous paragraph.

The above selection effect started from the earliest observations by
H.W. Babcock, who first identified strong magnetic fields in Ap stars after
many unsuccessful attempts. Indeed, stellar magnetic data sets now
available exhibit a strong tendency to present stars with strong or even
extreme fields. This selection effect can be avoided only when
measuring a ``canonical'' distribution of the magnetic fields for all stars
in a fixed volume of space.
We are aware that there exists general understanding of this problem,
and that there are observational projects of this type in progress.

The amount of necessary observational effort is very large, and it will
take years to complete. Our paper, however, was prepared taking
into account all existing $B_e$ measurements disregarding the
observational selection.

\section{Comments on CP classification }

Classification of chemically peculiar stars represents a very complex
problem. Commonly adopted criteria of classification rely on the presence
of particular elements or groups of elements in the spectra of Ap stars.
Such observables represent only the surface properties of the magnetic field
configuration of a star, and the resulting surface chemical anomalies.
The resulting classification of Ap stars into subclasses is very complex and
not unique, which is also reflected in Table 1 of this paper.

The referee suggested that since the existing classifications of
magnetic Ap stars are very inhomogeneous, one could divide them by
colour, $(U-B)_0$ for example. Such a choice would give a rough division
of Ap stars collected here by mass, which may be a more physically meaningful
parameter than the surface peculiarities.

We agree that one should seek for classification criteria among Ap stars
which are more physically meaningful than just the apparent surface
peculiarities. This will be a subject of our research in the near future.
In this paper, however, we adopt spectroscopic classification of
chemical peculiarities in various CP stars.

\subsection{HgMn stars}

The group of HgMn stars exhibits rather inhomogeneous content, similar
th the Sr group discussed in previous Sections (cf. Figs 2--3). There exists
small group of classical HgMn stars (e.g. $\iota$ CrB and $\alpha$ And)
for which no really convincing evidence of longitudinal fields is
available. There exist also other Ap stars (such as HD 21699 and 79158)
which display Hg or Mn along with numerous other peculiarities in their
spectra. These subgroups should be investigated separately.

The most actual list of both all HgMn stars and classical HgMn stars (the
latter are objects with pure HgMn peculiarity) has been recently published 
in Adelman et al. (2003). 

In the case of HgMn stars we have investigated the distribution functions $N_{Int}$
for the whole the group (see Fig. 8 and Table 5), and for only the classical
HgMn stars (Fig. 9 and Table 6). Fig. 9 clearly shows the well-known
fact that the classical HgMn stars have very weak longitudinal magnetic
fields. They are substantially different than other Hg or Mn stars, which
simultaneously exhibit also other chemical peculiarities. The latter
stars exhibit sometimes strong fields $\langle B_e \rangle$.

Fig. 9 shows that only three classical HgMn stars apparently exhibit noticeable
magnetic field $\langle B_e \rangle$: HD 172044, HD 210873, and HD 221507.
However, in all three cases the accuracy of $B_e$ observations is relatively
low. We speculate that their $\langle B_e \rangle$ reflect essentially errors
of measurement, and that high precision $B_e$ measurements will yield
much weaker averaged longitudinal magnetic fields for all three HgMn
stars.

One should keep in mind that the detailed investigation of various subclasses
of chemical peculiarities among CP stars is limited by the small number of
stars in subclaasses. For example. there are only 15 classical HgMn stars
in our compilation with which to construct Fig. 9 and Table 6.

   \begin{figure}
    \resizebox{\hsize}{!}{\includegraphics{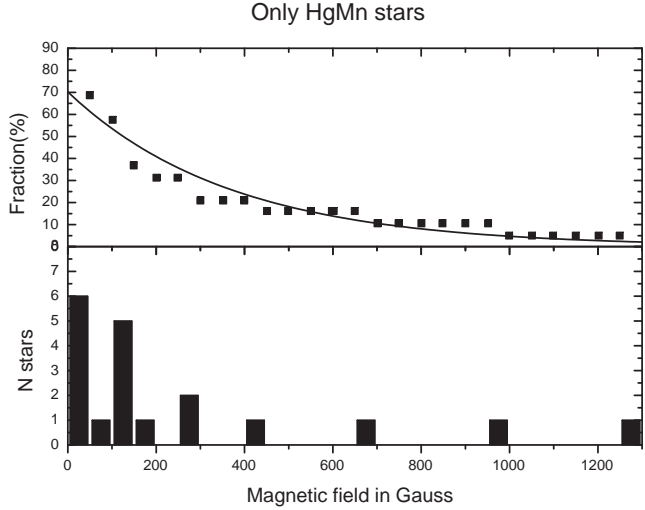}}
    \caption[]{The same for classical HgMn stars.  }
    \label{fig:fig8}
   \end{figure}

\section{Summary and conclusions }

The most important results of this paper can be summarized in the
following list:

{\bf 1.}
We present an extensive list of the averaged quadratic effective magnetic
fields $\langle B_e \rangle$ for main sequence and giant stars. Individual
$B_e$ observations were compiled from the existing literature,
and were further processed to obtain a homogeneous set of averaged effective
magnetic fields. We consider our averaged values of $\langle B_e \rangle$
as a reasonable  representative measure of the field strength in the atmosphere of
a given star. This is because the value of $\langle B_e \rangle$ results
directly from the observed effective magnetic field strengths $B_e$ and
is a strictly model-independent quantity.

Moreover, it is a single scalar parameter which describes the magnetic field
of a star even if the number of individual $B_e$ is low or the $B_e$
observations are noisy. In such a case the full curve describing $B_e(\phi)$
changes with rotational phase $\phi$ cannot be constructed.

{\bf 2.}
We have determined for the first time that the relation between the number
of occurences $N_{Int}$ of the magnetic field higher than a specified
$\langle B_e \rangle$ is given by the decreasing exponential function, at
least starting from the minimum value of $\langle B_e \rangle \approx 100$ G
\begin{equation}
N_{Int} (\langle B_e \rangle) = N_{tot} \, {a_1 \over {100 \%}}\,
  \exp \, (-\langle B_e \rangle /a_2) \, .
\end{equation}
Therefore the number distribution function $N(\langle B_e \rangle )$
of Ap type stars is also given by a decreasing exponential function.
This relations is found to hold for all analysed subclasses: Am, Si, He-weak,
He-rich, HgMn, SrCrEu, and all Sr type stars. We determined and listed
values of the parameters $a_1$ and $a_2$ for each subclass, see Table 1.


{\bf 3.}
We cannot rule out the possibility, that this exponential relation represents
just the tail of the true distribution, with its maximum hidden below
$\langle B_e \rangle \approx 100$ G. This is because our Figures and
fitting curves have limited resolution in the independent variable
$\langle B_e \rangle $, which is limited by the observational errors and limited sample sizes to the
width of the average bin, typically of the order of 100 G (in each individual sample
the value of the resolution is between 25 G and 200 G).

{\bf 4.} Our results demonstrate that the number distribution of the
averaged quadratic effective magnetic fields $N (\langle B_e \rangle)$
is not similar in any way to tail of the Gaussian distribution,
which would be proportional to $\exp (-\langle B_e \rangle ^2/a_2^2)$.

\vskip 7 mm  \noindent
The analysis presented in this paper is concentrated on the integrated
distribution function $N_{Int} (B)$, due to the rather low number of stars
available in most chemical peculiarity classes. Still, the function
$N_{Int}$ seems relatively smooth in all the subclasses, and is credibly
represented by an exponent. However, some small distortions can be easily
seen in upper panels of all the Figures 1--9.

The distribution function $N(B)$ is the first derivative of $N_{Int}$,
and obviously all numerical distortions of the latter involve
fluctuations of the derivative. This is seen in lower panels of Figures
1--9, in which directly measured distribution functions exhibit serious
noise. Therefore the exponential shape of the distribution function $N(B)$,
given in Eq. (8), is just an  extrapolation of the smoothed $N_{Int}$,
which is not inconsistent with the measured $N(B)$.

We exclude from the above rule region of the lowest magnetic fields $B$ of
extend comparable with the resolution ${\delt } B$ of our histograms.

\begin{acknowledgements}

We are grateful to J.D. Landstreet, the referee, for his criticism
and numerous suggestions which helped us to improve this paper.
Our research is based on data compiled and posted in the SIMBAD, ADS,
and CDS databases.
We acknowledge support from the Polish Committee for Scientific
Research grant No. 2 P03D 021 22.

\end{acknowledgements}

\par
\newpage


\begin{table}
\caption{List of He-weak stars. }
\label{tab:col}
\renewcommand{\arraystretch}{1.1}
\begin{tabular}{|r|r|c|l|}
\hline
 HD &	HR  & Name &  Sp. type \\
\hline
  4778 & 234  & GO And	       &    A1 CrSrEu He-w       \\[-1.0pt]
  5737 & 280  &$\alpha$ Scl    &    B6p SrTi He-w        \\[-1.0pt]
 19400 & 939  &$\theta$ Hyi    &    B3V+A0IV He-w        \\[-1.0pt]
 19805 &      &PPM45935        &    A0 He-w Si           \\[-1.0pt]
 21699 & 1063 & V396 Per       &    B8IIIp MnSi He-w     \\[-1.0pt]
 22470 & 1100 & 20 Eri	       &    B9p Si He-w          \\[-1.0pt]
 22920 & 1121 & 22 Eri	       &    B8p Si4200 He-w      \\
\hline
\end{tabular}
\end{table}

\par
\medskip

\begin{table}
\caption{List of He-rich stars. }
\label{tab:col}
\renewcommand{\arraystretch}{1.1}
\begin{tabular}{|r|r|c|l|}
\hline
 HD &	HR  & Name &  Sp. type \\
\hline
 35912& 1820&			 &B2V He-r	\\[-1.0pt]
 36430& 1848&			 &B2V He-r	\\[-1.0pt]
 36485& 1851&$\delta$ Ori B      &B2p He-r      \\[-1.0pt]
 37017& 1890&			 &B1.5V He-r	\\[-1.0pt]
 37479& 1932&$\sigma$ Ori E      &B2Vp He-r Be  \\[-1.0pt]
 37776&     &PPM175998           &B3 He-r       \\[-1.0pt]
 47777&     &PPM151028           &B2 He-r       \\[-1.0pt]
 58260&     &PPM283900           &B3 He-r       \\[-1.0pt]
 60344&     &PPM252646           &B3 He-r       \\[-1.0pt]
 64740& 3089&			 &B1.5Vp He-r	\\[-1.0pt]
 93030& 4199&$\theta$ Car        &B0Vp He-r     \\[-1.0pt]
 96446&     &PPM339754           &B2 He-r       \\[-1.0pt]
120640& 5206&			 &B2Vp He-r	\\[-1.0pt]
133518&     &PPM343417           &B3 He-r       \\[-1.0pt]
151346&     &PPM265974           &B4 He-r       \\[-1.0pt]
177003& 7210&			 &B2.5IV He-r	\\[-1.0pt]
184927&     &PPM83182            &B2 He-r       \\[-1.0pt]
186205&     &PPM168470           &B3 He-r       \\[-1.0pt]
209339& 8399&			 &B0IV He-r	\\
\hline
\end{tabular}
\end{table}

\par
\medskip

\begin{table}
\caption{List of Am-stars. }
\label{tab:col}
\renewcommand{\arraystretch}{1.1}
\begin{tabular}{|r|r|c|l|}
\hline
 HD &	HR  & Name &  Sp. type \\
\hline
 20210& 976 & V423 Per          & A1m              \\[-1.0pt]
 27962& 1389& 68 Tau		& A3V Am	   \\[-1.0pt]
 29140& 1458& 88 Tau		& A5 Am 	   \\[-1.0pt]
 29173& 1460&                   & A1m              \\[-1.0pt]
 31295& 1570&$\pi^1$ Ori        & A0V Am           \\[-1.0pt]
 48915& 2491&$\alpha$ CMa       & A1Vm             \\[-1.0pt]
 56495&     &PPM190112          & Am               \\
\hline
\end{tabular}
\end{table}

\par
\medskip

\begin{table}
\caption{List of Mn and Hg stars. }
\label{tab:col}
\renewcommand{\arraystretch}{1.1}
\begin{tabular}{|r|r|c|l|}
\hline
 HD &	HR  & Name &  Sp. type \\
\hline
   358& 15  &$\alpha$ And       & B9p HgMn                \\[-1.0pt]
 21699& 1063& V396 Per		& B8IIIp MnSi He-w	  \\[-1.0pt]
 22316& 1094& GC 4315           & B9p HgMnCrSi            \\[-1.0pt]
 23408& 1149& 20 Tau		& B8IIIP Hg He-w	  \\[-1.0pt]
 27295& 1339& 53 Tau            & B9HgMn                  \\[-1.0pt]
 27376& 1347&41 $\upsilon^4$ Eri& B9V HgMn                \\[-1.0pt]
 33904& 1702&$\mu$ Lep          & B9p HgMn                \\
\hline
\end{tabular}
\end{table}

\begin{table}
\caption{List of only HgMn stars. }
\label{tab:col}
\renewcommand{\arraystretch}{1.1}
\begin{tabular}{|r|r|c|l|}
\hline
 HD &	HR  & Name &  Sp. type \\
\hline
   358&   15&$\alpha$ And       & B9p HgMn                \\[-1.0pt]
 27295& 1339& 53 Tau            & B9HgMn                  \\[-1.0pt]
 27376& 1347&41 $\upsilon^4$ Eri& B9V HgMn                \\[-1.0pt]
 33904& 1702&$\mu$ Lep          & B9p HgMn                \\[-1.0pt]
 37752& 1951&			& B8p HgMn		  \\[-1.0pt]
 63975& 3059&$\zeta$ CMi        & B8HgMn                  \\[-1.0pt]
 75333& 3500& 14 Hya            & B9HgMn                  \\[-1.0pt]
 78316& 3623&$\kappa$ Cnc       & B8IIIp MnHg             \\[-1.0pt]
120709& 5210& 3 Cen A           &      HgMn               \\[-1.0pt]
129174& 5475&29 $\pi^1$ Boo     & B9p HgMn                \\[-1.0pt]
141556& 5883&5 $\chi$ Lup       & B9IV HgMn               \\[-1.0pt]
143807& 5971&14 $\iota$ CrB     & AOp MnHg                \\[-1.0pt]
144206& 5982&6 $\nu$ Her        & B9III MnHg              \\[-1.0pt]
145389& 6023& 11 $\phi$ Her     & B9p HgMn                \\[-1.0pt]
172044& 6997&                   &B8II-IIIp HgMn           \\[-1.0pt]
172883& 7028&GC 25559           &B9p SiHgMn               \\[-1.0pt]
210873& 8473&                   &B9p HgMn                 \\[-1.0pt]
216494& 8704&74 HI Aqr          &B9III HgMn  binary star  \\[-1.0pt]
221507& 8936&$\beta$ Scl        &B9.5IVp HgMnEu           \\
\hline
\end{tabular}
\end{table}

\par
\medskip
\begin{table}
\caption{List of Si stars }
\label{tab:col}
\renewcommand{\arraystretch}{1.1}
\begin{tabular}{|r|r|c|l|}
\hline
 HD &	HR  & Name &  Sp. type \\
\hline
  8855&     & BD+43 301         & B9 SiCr                  \\[-1.0pt]
  9996&  465& GY And		& B9p CrEuSi		   \\[-1.0pt]
 10221&  478& 43 Cas		& A0p SiSrCr		   \\[-1.0pt]
 10783&     & UZ Psc		& A2 SrCrSi		   \\[-1.0pt]
 11187&     & BD+54 393         & A0 SiCrSr                \\[-1.0pt]
 11503&  545& $\gamma^2$ Ari    & A1p SiCrEu               \\[-1.0pt]
 12288&     & V540 Cas		& A2p CrSi		   \\
\hline
\end{tabular}
\end{table}

\par
\medskip

\begin{table}
\setcounter{table}{6}
\caption{List of Si stars -- continued }
\label{tab:col}
\renewcommand{\arraystretch}{1.1}
\begin{tabular}{|r|r|c|l|}
\hline
 HD &	HR  & Name &  Sp. type \\
\hline
 66255& 3151& PY Pup		& A0p Si		   \\[-1.0pt]
 68351& 3215& 15 Cnc		& A0p SiCrSr		   \\[-1.0pt]
 70331&     & CoD-47 3803	& B8p Si		   \\[-1.0pt]
 71866&     & TZ Lyn		& A1 SiSrEu		   \\[-1.0pt]
 73340& 3413& HV Vel		& B9 Si 		   \\[-1.0pt]
 74521& 3465& 49 Cnc		& A1p EuSiCr		   \\[-1.0pt]
 77350& 3595&69 $\nu$ Cnc       & B9p Si4012,CrHg          \\
\hline
\end{tabular}
\end{table}

\par
\medskip

\begin{table}
\setcounter{table}{6}
\caption{List of Si stars -- continued }
\label{tab:col}
\renewcommand{\arraystretch}{1.1}
\begin{tabular}{|r|r|c|l|}
\hline
 HD &	HR  & Name &  Sp. type \\
\hline
147010&     & V933 Sco          & B9p SiCrFeSr             \\[-1.0pt]
147550& 6096&			& B9V Si		   \\[-1.0pt]
147890&     & V936 Sco		& B9.5p SiSr		   \\[-1.0pt]
148199&     & GC 22126		& B9 SiSr He-w		   \\[-1.0pt]
148330& 6127& DQ Dra		& A2 Si3955v:SrCrEu	   \\[-1.0pt]
149822& 6176& V773 Her		& B9p SiCrSr		   \\[-1.0pt]
149911& 6179& GC 22360          & A0p CrEuSiSr             \\
\hline
\end{tabular}
\end{table}

\par
\medskip

\begin{table}
\caption{List of SrCrEu stars. }
\label{tab:col}
\renewcommand{\arraystretch}{1.1}
\begin{tabular}{|r|r|c|l|}
    \hline
 HD &	HR  & Name &  Sp. type \\
\hline
  2453&     & GR And		& A1 SrEuCr		\\[-1.0pt]
  3980&  183&$\xi$ Phe          & A7 SrCrEu             \\[-1.0pt]
  4778&  234& GO And		& A1 CrSrEu He-w	\\[-1.0pt]
  5797&     & V551 Cas		& A0 SrCrEu		\\[-1.0pt]
  6532&     & AP Scl		& A3p SrCr		\\[-1.0pt]
  8441&     & HN And		& A2 SrCrEu		\\[-1.0pt]
  9996&  465& GY And		& B9p CrEuSi		\\
\hline
\end{tabular}
\end{table}

\par
\medskip

\begin{table}
\setcounter{table}{7}
\caption{List of SrCrEu stars - continued}
\label{tab:col}
\renewcommand{\arraystretch}{1.1}
\begin{tabular}{|r|r|c|l|}
    \hline
 HD &	HR  & Name &  Sp. type \\
\hline
103192& 4552&$\beta$ Hya        & B9IIIp SiCrSr         \\[-1.0pt]
103498& 4561& 65 UMa		& A1p SrCrEu		\\[-1.0pt]
108662& 4752& 17 Com A		& A0p SrCrEuSi		\\[-1.0pt]
108945& 4766&        UU Com     & A2pv SrCr             \\[-1.0pt]
110066& 4816& AX CVn            & A1(0)p SrCrEu         \\[-1.0pt]
111133& 4854& EP Vir		& A1p SrCrEu		\\[-1.0pt]
112185& 4905&77 $\epsilon$ UMa  & A1 CrEuMn             \\
\hline
\end{tabular}
\end{table}

\par
\medskip

\begin{table}
\setcounter{table}{7}
\caption{List of SrCrEu stars - continued }
\label{tab:col}
\renewcommand{\arraystretch}{1.1}
\begin{tabular}{|r|r|c|l|}
    \hline
 HD &	HR  & Name &  Sp. type \\
\hline
184905&     & V1264 Cyg         & A0 SiSrCr              \\[-1.0pt]
187474& 7552& V3961 Sgr         & A0p EuCrSi             \\[-1.0pt]
188041& 7575& V1291 Aql 	& A6p SrCrEu	         \\[-1.0pt]
191742&     & PPM 59411         & A5 SrCrEu              \\[-1.0pt]
193756&     & CPD-52 11681	& A9p SrCrEu	         \\[-1.0pt]
196502& 7879& 73 AF Dra 	& A2p SrCrEu             \\[-1.0pt]
200177&     & PPM 60816         & A1 CrSrEu              \\[-1.0pt]
201601& 8097&5 $\gamma$ Equ     & A9p SrCrEu             \\[-1.0pt]
203006& 8151&$\theta^1$ Mic     & A2 MgCrSrEu            \\[-1.0pt]
203932&     & CoD-30 18600	& A5p SrEu		 \\[-1.0pt]
204131& 8206& GC 30005		& B9p SiSrCr		 \\[-1.0pt]
205087& 8240&			& B9p SiSrCrEu		 \\[-1.0pt]
208095& 8357&			& B6IV-V SiSrCrEu	 \\[-1.0pt]
212454& 8535&			& B8III SrCrEuHg He-w    \\[-1.0pt]
216018&     & BD-12 6357	& A7p SrCrEu		 \\[-1.0pt]
216533&     & MX Cep		& A1 SrCr		 \\[-1.0pt]
217522&     & CoD-45 14901	& A5p SrEuCr		 \\[-1.0pt]
218495&     & CPD-64 4322	& A2p EuSr		 \\[-1.0pt]
220825& 8911&8 $\kappa$ Psc     & A1p CrSi:SrEu          \\[-1.0pt]
221394& 8933& GC 32719		& A0p SrCrEu:Si 	 \\[-1.0pt]
221568&     & V436 Cas		& A1 SrCrEu		 \\[-1.0pt]
221760& 8949& $\iota$ Phe       & A2Vp SrCrEuSi          \\[-1.0pt]
223640& 9031& ET Aqr		& B9p SiSrCr		 \\[-1.0pt]
224801& 9080& CG And		& B9.5p SiCrEu		 \\[-1.0pt]
335238&     & BD+29 4202	& A1p CrEu		 \\
\hline
\end{tabular}
\end{table}

\begin{table}
\caption{List of other Sr stars. }
\label{tab:col}
\renewcommand{\arraystretch}{1.1}
\begin{tabular}{|r|r|c|l|}
    \hline
 HD &	HR  & Name &  Sp. type \\
\hline
  5737&  280&$\alpha$ Scl       & B6p SrTi He-w          \\[-1.0pt]
  8855&     & PPM 44016         & B9 SiCr                \\[-1.0pt]
 12288&     & V540 Cas		& A2p CrSi		 \\[-1.0pt]
 22316& 1094& GC4315		& B9p HgMnCrSi		 \\[-1.0pt]
 22407&     & PPM 185972        & Cr                     \\[-1.0pt]
 23387&     & PPM 92834         & A0 CrSi He-w           \\[-1.0pt]
 27309& 1341& 56 Tau		& A0p SiCr		 \\
\hline
\end{tabular}
\end{table}

\par

\appendix
\section{The catalogue of averaged effective magnetic fields}

Table {\bf A.1} presents the complete listing of stars with individual averaged
quadratic values of $\langle B_e \rangle$ and the additional data. Columns
of the Table list: HD number, spectral type of the star, number $N$ of
individual $B_e$ values, standard deviation $\sigma$, corresponding value
of $\chi ^2$ per one degree of freedom, method of $B_e$ determination, and
reference numbers. The list of references for Table 1 is given at the end
of this Appendix.

\newdimen\digitwidth
\setbox0=\hbox{\rm0}
\digitwidth=\wd0
\catcode`?=\active
\def?{\kern\digitwidth}

%
%


\par
\medskip
\clearpage
\newpage

\newcommand{\di}{\displaystyle}

\begin{table}
\caption{List of stellar magnetic fields }
\renewcommand{\arraystretch}{1.1}
\begin{tabular}{|r|l|r|r|r|r|l|l|}
    \hline
 HD \hskip1mm & Sp.type &N &$\langle B_e \rangle \hskip1mm$ &
    $ \sigma\, \langle B_e \rangle$ &
    $\chi^2 \hskip1mm$ & Method & References/notes      \\
\hline
   358& B9p   &  14&      65.3&    134.5&      1.45&  all &   2   56  127  140 \\[-1.8pt]
   358&       &  13&      67.8&     67.7&      1.56&  all &   2  127  140      \\[-1.8pt]
   358&       &   7&      57.3&     46.8&      1.81&  Hl  &   2  127           \\[-1.8pt]
   358&       &   6&      78.2&     85.9&      1.27&  Met & 140                \\[-1.8pt]
   432& F2IV  &   3&     220.1&    175.6&      1.11&  Hl  & 235  266           \\[-1.8pt]
   886& B2IV  &  13&     670.3&    357.6&      2.16&  all &  77   96  232      \\[-1.8pt]
   886&       &  12&     694.5&    368.9&      2.18&  Hl  &  77   96           \\
\hline
\end{tabular}
\end{table}

\par
\medskip
\clearpage
\newpage
\begin{table}
\setcounter{table}{0}
\caption{List of stellar magnetic fields -- continued}
\renewcommand{\arraystretch}{1.1}
\begin{tabular}{|r|l|r|r|r|r|l|l|}
    \hline
 HD \hskip1mm & Sp.type &N &$\langle B_e \rangle \hskip1mm$ &
    $ \sigma\, \langle B_e \rangle$ &
    $\chi^2 \hskip1mm$ & Method & References/notes	\\
\hline
 18296&       &  51&     503.6&    227.6&      4.72&  Met &   1   14  327      \\[-1.8pt]
 18296&       &  14&     191.0&    177.5&      1.29&  Hl  &   2  197  243      \\[-1.8pt]
 19356& B8V   &   1&        .0&    320.0&       .00&  Met &  56                \\[-1.8pt]
 19400& B3V+A0&   4&     206.7&    156.9&      1.61&  Hl  &  37                \\[-1.8pt]
 19445& A4p   &   1&     415.0&    550.0&       .57&  Met &   1                \\[-1.8pt]
 19653& A0    &   1&     548.0&    115.0&     22.71&  Met & 327                \\[-1.8pt]
 19805& A0V   &   4&    1039.6&    239.6&     17.28&  Met & 327                \\
\hline
\end{tabular}
\end{table}

\par
\medskip
\clearpage
\newpage
\begin{table}
\setcounter{table}{0}
\caption{List of stellar magnetic fields -- continued}
\renewcommand{\arraystretch}{1.1}
\begin{tabular}{|r|l|r|r|r|r|l|l|}
    \hline
 HD \hskip1mm & Sp.type &N &$\langle B_e \rangle \hskip1mm$ &
    $ \sigma\, \langle B_e \rangle$ &
    $\chi^2 \hskip1mm$ & Method & References/notes	\\
\hline
 31648& A2    &   1&     180.0&    300.0&       .36&  Met & 286                \\[-1.8pt]
 32549& A0p   &   4&      50.2&    132.9&       .14&  Hl  &   2                \\[-1.8pt]
 32630& B3V   &   1&    1672.0&   1134.0&      2.17&  Met & 327                \\[-1.8pt]
 32633& B9p   & 120&    2760.6&    262.9&    171.85&  all &   1    2    8  175 \\[-1.8pt]
      &       &    &          &         &          &      & 184  285  310  327 \\[-1.8pt]
 32633&       &  88&    2857.4&    223.3&    180.96&  Met &   1    8  175  184 \\[-1.8pt]
      &       &    &          &         &          &      & 285  310  327      \\
\hline
\end{tabular}
\end{table}

\par
\medskip
\clearpage
\newpage
\begin{table}
\setcounter{table}{0}
\caption{List of stellar magnetic fields -- continued}
\renewcommand{\arraystretch}{1.1}
\begin{tabular}{|r|l|r|r|r|r|l|l|}
    \hline
 HD \hskip1mm & Sp.type &N &$\langle B_e \rangle \hskip1mm$ &
    $ \sigma\, \langle B_e \rangle$ &
    $\chi^2 \hskip1mm$ & Method & References/notes	\\
\hline
 40312& A0p   &  26&	 223.2&     57.3&     16.36&  Hl  &   2   60 - only $H_\beta$  \\[-1.8pt]
 40312&       &   9&	 340.6&    290.3&      1.58&  Hl  &  25      - only $H_\alpha$ \\[-1.8pt]
 40312&       &  10&    1025.8&    505.7&      6.66&  Met & 226                \\[-1.8pt]
 40312&       &  11&     213.2&     41.9&     48.09&  LSD & 310                \\[-1.8pt]
 41753& B3V   &   1&     120.0&    420.0&       .08&  LSD & 333                \\[-1.8pt]
 42474& M1ab  &   8&     528.1&    206.3&     38.56&  Met &   1   72           \\[-1.8pt]
 42616& A1    &   4&     620.4&    238.0&      6.80&  Met &   1                \\
\hline
\end{tabular}
\end{table}

\par
\medskip
\clearpage
\newpage
\begin{table}
\setcounter{table}{0}
\caption{List of stellar magnetic fields -- continued}
\renewcommand{\arraystretch}{1.1}
\begin{tabular}{|r|l|r|r|r|r|l|l|}
    \hline
 HD \hskip1mm & Sp.type &N &$\langle B_e \rangle \hskip1mm$ &
    $ \sigma\, \langle B_e \rangle$ &
    $\chi^2 \hskip1mm$ & Method & References/notes	\\
\hline
 62509& K0IIIb&  12&       6.6&     13.8&      1.43&  Met &  56  299  321  322 \\[-1.8pt]
 63401& B8III &   1&     280.0&    210.0&      1.78&  Hl  & 230                \\[-1.8pt]
 63975& B8    &   1&     100.0&    110.0&       .83&  LSD & 333                \\[-1.8pt]
 64486& B9p   &   6&     854.7&    513.6&      5.40&  all &  91  230           \\[-1.8pt]
 64486&       &   3&    1118.3&    665.2&      7.94&  Met &  91                \\[-1.8pt]
 64486&       &   3&     458.5&    291.8&      2.87&  Hl  & 230                \\[-1.8pt]
 64740& B1.5Vp&  21&     571.9&    113.8&     24.99&  all &  24  135           \\
\hline
\end{tabular}
\end{table}

\par
\medskip
\clearpage
\newpage
\begin{table}
\setcounter{table}{0}
\caption{List of stellar magnetic fields -- continued}
\renewcommand{\arraystretch}{1.1}
\begin{tabular}{|r|l|r|r|r|r|l|l|}
    \hline
 HD \hskip1mm & Sp.type &N &$\langle B_e \rangle \hskip1mm$ &
    $ \sigma\, \langle B_e \rangle$ &
    $\chi^2 \hskip1mm$ & Method & References/notes	\\
\hline
 83368& A8p   &  34&     576.8&    264.1&      7.16&  all & 184  186  256  285 \\[-1.8pt]
 83368&       &  25&     604.8&    290.7&      6.65&  Met & 184  256  285      \\[-1.8pt]
 83368&       &   9&     490.8&    169.5&      8.56&  Hl  & 186                \\[-1.8pt]
 83625& A0    &   2&     306.1&    405.2&      1.18&  Hl  & 230                \\[-1.8pt]
 84367& F7V   &   1&     210.0&     80.0&      6.89&  Met & 120                \\[-1.8pt]
 84441& G0II  &   3&      29.6&     23.3&     22.44&  Met &  56  322           \\[-1.8pt]
 84999& F2IV  &   6&     376.2&    466.5&       .59&  Hl  & 235  266           \\
\hline
\end{tabular}
\end{table}

\par
\medskip
\clearpage
\newpage
\begin{table}
\setcounter{table}{0}
\caption{List of stellar magnetic fields -- continued}
\renewcommand{\arraystretch}{1.1}
\begin{tabular}{|r|l|r|r|r|r|l|l|}
    \hline
 HD \hskip1mm & Sp.type &N &$\langle B_e \rangle \hskip1mm$ &
    $ \sigma\, \langle B_e \rangle$ &
    $\chi^2 \hskip1mm$ & Method & References/notes	\\
\hline
102870& F8V   &   1&     270.0&    151.0&      3.20&  Met & 124                \\[-1.8pt]
102870&       &   2&       7.4&     18.5&       .15&  Met & 334                \\[-1.8pt]
103192& B9IIIp&   5&     204.2&    104.1&      3.92&  Hl  &   2  230           \\[-1.8pt]
103287& A0Ve  &   1&      25.0&     50.0&       .25&  Hl  &  77                \\[-1.8pt]
103287&       &   6&     235.2&    221.4&       .98&  all &  77  333  334      \\[-1.8pt]
103498& A1p   &   6&     333.0&    138.7&      6.24&  Met &  41  327           \\[-1.8pt]
104321& A5V   &   1&      93.0&     53.0&      3.08&  LSD & 333                \\
\hline
\end{tabular}
\end{table}

\par
\medskip
\clearpage
\newpage
\begin{table}
\setcounter{table}{0}
\caption{List of stellar magnetic fields -- continued}
\renewcommand{\arraystretch}{1.1}
\begin{tabular}{|r|l|r|r|r|r|l|l|}
    \hline
 HD \hskip1mm & Sp.type &N &$\langle B_e \rangle \hskip1mm$ &
    $ \sigma\, \langle B_e \rangle$ &
    $\chi^2 \hskip1mm$ & Method & References/notes	\\
\hline
115708& A3    &  14&     927.1&    405.8&      6.53&  all &   1  248           \\[-1.8pt]
115708&       &  11&    1001.0&    440.3&      6.72&  Hl  & 248                \\[-1.8pt]
115735& A0V   &   1&      10.0&    380.0&       .00&  LSD & 333                \\[-1.8pt]
116114& F0p   &   1&    1923.0&    113.0&          &  Met & 256                \\[-1.8pt]
116458& A0p   &  37&    1925.5&    273.0&    242.81&  Met &  93  184  187  256 \\[-1.8pt]
      &       &    &          &         &          &      & 285                \\[-1.8pt]
116656& A1V   &   5&      89.8&     73.2&       .94&  Met & 333 334 $\zeta$ UMa A\\
\hline
\end{tabular}
\end{table}

\par
\medskip
\clearpage
\newpage
\begin{table}
\setcounter{table}{0}
\caption{List of stellar magnetic fields -- continued}
\renewcommand{\arraystretch}{1.1}
\begin{tabular}{|r|l|r|r|r|r|l|l|}
    \hline
 HD \hskip1mm & Sp.type &N &$\langle B_e \rangle \hskip1mm$ &
    $ \sigma\, \langle B_e \rangle$ &
    $\chi^2 \hskip1mm$ & Method & References/notes	\\
\hline
127972& B2IV  &   1&      66.0&     76.0&       .75&  Hl  & 200                \\[-1.8pt]
128167& F2V   &   2&       9.6&     35.7&       .13&  Hl  &  77                \\[-1.8pt]
128167& F2V   &   1&       8.0&     13.0&       .38&  LSD & 333                \\[-1.8pt]
128220& G     &   3&     371.9&    711.4&       .36&  all & 232  275           \\[-1.8pt]
128775& B9p   &   1&     435.0&    185.0&      5.53&  Hl  & 168                \\[-1.8pt]
128898& A0p   &  31&     644.6&    324.3&      6.55&  all &   2   81   93  184 \\[-1.8pt]
      &       &    &          &         &          &      & 256  285           \\
\hline
\end{tabular}
\end{table}

\par
\medskip
\clearpage
\newpage
\begin{table}
\setcounter{table}{0}
\caption{List of stellar magnetic fields -- continued}
\renewcommand{\arraystretch}{1.1}
\begin{tabular}{|r|l|r|r|r|r|l|l|}
    \hline
 HD \hskip1mm & Sp.type &N &$\langle B_e \rangle \hskip1mm$ &
    $ \sigma\, \langle B_e \rangle$ &
    $\chi^2 \hskip1mm$ & Method & References/notes	\\
\hline
137949& F0p   &  18&    1497.7&    197.0&     63.97&  Met &   1   26   91  256 \\[-1.8pt]
138749& B6III &   1&     100.0&    190.0&       .28&  Hl  & 200                \\[-1.8pt]
139365& B2.5V &   2&      47.4&     70.1&       .49&  Hl  & 168                \\[-1.8pt]
139525& B8    &   4&     214.8&    234.1&       .88&  Hl  & 168                \\[-1.8pt]
140160& A0p   &   9&     859.1&    712.3&      1.51&  Hl  &   2   25           \\[-1.8pt]
140160&       &   1&     230.0&    120.0&      3.67&  LSD & 333                \\[-1.8pt]
140573& K2IIIb&   1&       2.4&      1.8&       .25&  Met & 321                \\
\hline
\end{tabular}
\end{table}

\par
\medskip
\clearpage
\newpage
\begin{table}
\setcounter{table}{0}
\caption{List of stellar magnetic fields -- continued}
\renewcommand{\arraystretch}{1.1}
\begin{tabular}{|r|l|r|r|r|r|l|l|}
    \hline
 HD \hskip1mm & Sp.type &N &$\langle B_e \rangle \hskip1mm$ &
    $ \sigma\, \langle B_e \rangle$ &
    $\chi^2 \hskip1mm$ & Method & References/notes	\\
\hline
147394& B5IV  &   1&     230.0&    190.0&      1.47&  Hl  &  77                \\[-1.8pt]
147394&       &   1&      33.0&     87.0&       .14&  LSD & 333                \\[-1.8pt]
147550& B9V   &   4&     416.5&    430.2&      1.29&  Met &  91                \\[-1.8pt]
147888& B5V   &   1&     100.0&    200.0&       .25&  Hl  & 168                \\[-1.8pt]
147890& B9.5p &   4&     235.1&    256.4&       .85&  Hl  & 168                \\[-1.8pt]
148112& A0p   &  22&     649.7&    441.1&      2.10&  all &   2   25  327      \\[-1.8pt]
148112&       &  20&     677.8&    436.1&      2.28&  Hl  &   2   25  327      \\
\hline
\end{tabular}
\end{table}

\par
\medskip
\clearpage
\newpage
\begin{table}
\setcounter{table}{0}
\caption{List of stellar magnetic fields -- continued}
\renewcommand{\arraystretch}{1.1}
\begin{tabular}{|r|l|r|r|r|r|l|l|}
    \hline
 HD \hskip1mm & Sp.type &N &$\langle B_e \rangle \hskip1mm$ &
    $ \sigma\, \langle B_e \rangle$ &
    $\chi^2 \hskip1mm$ & Method & References/notes	\\
\hline
167817& G0V   &   4&     283.6&    131.0&      4.01&  Met & 223                \\[-1.8pt]
168605& A0    &   1&    1784.0&    605.0&      8.70&  Hl  & 326                \\[-1.8pt]
168733& B8p   &  30&     815.4&    276.2&     19.58&  Met &  86   93  184  256 \\[-1.8pt]
      &       &    &          &         &          &      & 285                \\[-1.8pt]
168733&       &  20&     832.4&    305.3&     13.78&  Met &  86   93           \\[-1.8pt]
169022&B9.5III&   3&      70.5&     60.0&      1.38&  Hl  & 183                \\[-1.8pt]
169027& A0    &   1&     325.0&    820.0&       .16&  Met & 275                \\
\hline
\end{tabular}
\end{table}

\par
\medskip
\clearpage
\newpage
\begin{table}
\setcounter{table}{0}
\caption{List of stellar magnetic fields -- continued}
\renewcommand{\arraystretch}{1.1}
\begin{tabular}{|r|l|r|r|r|r|l|l|}
    \hline
 HD \hskip1mm & Sp.type &N &$\langle B_e \rangle \hskip1mm$ &
    $ \sigma\, \langle B_e \rangle$ &
    $\chi^2 \hskip1mm$ & Method & References/notes	\\
\hline
187013& F5V   &   1&      20.0&     19.0&      1.11&  Met & 299                \\[-1.8pt]
187474& A0p   &  56&    1488.0&    143.9&    125.50&  Met &   1  184  256  285 \\[-1.8pt]
187642& A7IV  &   2&      46.4&     55.2&       .60&  Hl  &  77                \\[-1.8pt]
187642&       &   6&     109.8&     89.0&      4.35&  all &  77  334           \\[-1.8pt]
187929& F6Iab &  40&      39.0&     29.7&      2.53&  Met &  76  299  335  336 \\[-1.8pt]
187929&       &  20&      54.5&     41.1&      2.37&  Met & 335                \\[-1.8pt]
187929&       &  14&       7.6&      4.4&      3.33&  LSD & 336                \\
\hline
\end{tabular}
\end{table}

\par
\medskip
\clearpage
\newpage
\begin{table}
\setcounter{table}{0}
\caption{List of stellar magnetic fields -- continued}
\renewcommand{\arraystretch}{1.1}
\begin{tabular}{|r|l|r|r|r|r|l|l|}
    \hline
 HD \hskip1mm & Sp.type &N &$\langle B_e \rangle \hskip1mm$ &
    $ \sigma\, \langle B_e \rangle$ &
    $\chi^2 \hskip1mm$ & Method & References/notes	\\
\hline
202850& B9Iab &   1&     110.0&    900.0&       .01&  Hl  & 163                \\[-1.8pt]
203006& A2    &   4&     344.7&    153.5&      3.01&  all &   1    2           \\[-1.8pt]
203006&       &   3&     132.5&    105.2&      1.71&  Hl  &   2                \\[-1.8pt]
203932& A5p   &   2&     250.0&    150.5&      2.70&  Met & 256                \\[-1.8pt]
204041& A2V   &   2&     527.9&    405.3&      1.73&  Hl  & 183                \\[-1.8pt]
204131& B9p   &   7&     788.9&    535.0&      2.30&  Met & 327                \\[-1.8pt]
205021& B1IV  &  10&    1034.5&   1539.9&      2.86&  Hl  &  96                \\
\hline
\end{tabular}
\end{table}

\par
\medskip
\clearpage
\newpage
\begin{table}
\setcounter{table}{0}
\caption{List of stellar magnetic fields -- continued}
\renewcommand{\arraystretch}{1.1}
\begin{tabular}{|r|l|r|r|r|r|l|l|}
    \hline
 HD \hskip1mm & Sp.type &N &$\langle B_e \rangle \hskip1mm$ &
    $ \sigma\, \langle B_e \rangle$ &
    $\chi^2 \hskip1mm$ & Method & References/notes	\\
\hline
217833& B9III &  23&    3648.7&    697.5&     36.59&  all &  40  230  233  326 \\[-1.8pt]
217833&       &  18&    4105.5&    768.2&     45.42&  Met &  40  233           \\[-1.8pt]
217833&       &   5&     748.9&    336.5&      4.78&  Hl  & 230  326           \\[-1.8pt]
218393& Bpe   &   3&    1820.7&    416.6&     23.57&  Met & 327                \\[-1.8pt]
218398& F0    &   2&     436.6&    418.4&       .98&  Met &  41                \\[-1.8pt]
218495& A2p   &   1&     606.0&    280.0&      4.68&  Met & 256                \\[-1.8pt]
219749& B9p   &  26&    1000.5&    412.1&      6.38&  Hl  & 198  230  327      \\
\hline
\end{tabular}
\end{table}

\par\medskip \clearpage

\noindent {\bf Cross-reference list : }
\vskip 1 mm

\noindent
1  Babcock H.W., 1958.			            \eol
2  Borra E.F. \& Landstreet J.D., 1980.             \eol
3  Babcock H.W., 1956.			            \eol
4  Babcock H.W., 1954.			            \eol
5  Babcock H.W., 1960.			            \eol
6  Preston G.W. \& Pyper D.M., 1965.	            \eol
7  Preston G.W., 1967.	                            \eol
8  Preston G.W. \& St\c epie\'n K., 1968a.          \eol
9  Preston G.W. \& St\c epie\'n K., 1968b.          \eol
10  Conti P.S., 1969.                               \eol
11  Preston G.W. et al., 1969.	                    \eol
12  Preston G.W., 1969a.                            \eol
13  Preston G.W., 1969b.                            \eol
14  Preston G.W., 1969c.                            \eol
15  Wolff S.C., 1969a.                              \eol
16  Wolff S.C., 1969b.                              \eol
17  Wolff S.C. \& Wolff R.J., 1970.                 \eol
18  Preston G.W., 1970.                             \eol
19  Preston G.W. \& Wolff S.C., 1970.	            \eol
20  Conti P.S., 1970a.			            \eol
21  Preston G.W., 1972. 	                    \eol
22  Wolff S.C. \& Bonsack W.K., 1972.               \eol
23  Wolff S.C. \& Wolff R.J., 1972.                 \eol
24  Borra E.F. \& Landstreet J.D., 1979.            \eol
25  Landstreet J.D. et al., 1975.                   \eol
26  Wolff S.C., 1975.			            \eol
27  Borra E.F. \& Vaughan A.H., 1978.	            \eol
28  Landstreet J.D. \& Borra E.F., 1978.            \eol
29  Jones T.J. et al., 1974.	                    \eol
30  Borra E.F. \& Landstreet J.D., 1978.	    \eol
31  Vogt S.S. et al., 1980.			    \eol
32  Bonsack W.K., 1976a.                            \eol
33  Borra E.F. \& Landstreet J.D., 1977.	    \eol
34  Borra E.F., 1981.				    \eol
35  Preston G.W. \& St\c epie\'n K., 1968c.         \eol
36  Preston G.W., 1971. 			    \eol
37  Borra E.F et al,, 1983.			    \eol
38  Chunakova N.M. et al., 1981a.                   \eol
39  Plachinda S.I., 1986.			    \eol
40  Glagolevskij Yu.V. \& Chunakova N.M., 1985.	    \eol
41  Glagolevskij Yu.V. et al., 1985a.               \eol
42  Rakos K.D. et al., 1977.			    \eol
43  Pyper D.M., 1969.				    \eol
44  Huchra J., 1972.				    \eol
45  Angel J.R.P. \& Landstreet J.D., 1970.	    \eol
46  Bonsack W.K, \& Pilachowski C.A., 1974.	    \eol
47  Brown D.N. \& Landstreet J.D., 1981.	    \eol
48  Scholz G., 1979.				    \eol
49  Scholz G., 1971b.				    \eol
50  Preston G.W., 1969d.			    \eol
51  Preston G.W., 1969e.			    \eol
52  Wolff R.J., Wolff S.C., 1976.		    \eol
53  Sargent W.L.W. et al., 1967.		    \eol
54  Borra E.F., 1975a.				    \eol
55  Conti P.S., 1970b.				    \eol
56  Borra E.F. \& Landstreet J.D., 1973a.	    \eol
57  Borra E.F. \& Landstreet J.D., 1973b.	    \eol
58  Kodaira K. \& Unno W., 1969.		    \eol
59  Borra E.F. \& Dworetsky M.M., 1973. 	    \eol
60  Landstreet J.D. \& Borra E.F., 1977.	    \eol
61  Boesgaard A.M., 1974.			    \eol
62  Bonsack W.K. et al., 1974.			    \eol
63  Borra E.F. \& Vaughan A.H., 1977.		    \eol
64  Borra E.F., 1980.				    \eol
65  Anderson C.M., Hartmann L.W., 1976. 	    \eol
66  Mullan D.J. \& Bell R.A., 1976.		    \eol
67  Kemp J.C. \& Wolstencroft R.D., 1973a.	    \eol
68  Marcy G.W., 1981.				    \eol
69  Kemp J.C. \& Wolstencroft R.D., 1973b.	    \eol
70  Vogt S.S., 1980.				    \eol
71  Wolff S.C., 1973.				    \eol
72  Slovak M.H., 1982.				    \eol
73  Borra E.F. \& Vaughan A.H., 1976.		    \eol
74  Angel J.R.P. et al., 1973.			    \eol
75  Borra E.F., 1975b.				    \eol
76  Borra E.F. et al., 1981.			    \eol
77  Landstreet J.D., 1982.			    \eol
78  Robinson R.D. \& Worden S.P., 1980. 	    \eol
79  Barker P.K. et al., 1981.			    \eol
80  Bonsack W.K., 1976b.                            \eol
81  Borra E.F. \& Landstreet J.D., 1975.	    \eol
82  Boesgaard A.M. et al., 1975.		    \eol
83  Wolff S.C., Preston G.W., 1978.		    \eol
84  Wolff S.C., 1978.				    \eol
85  Wolff S.C. \& Hagen W., 1976.		    \eol
86  Jones T.J., Wolff S.C., 1974.		    \eol
87  Bonsack W.K., 1981. 			    \eol
88  Hockey M.S., 1969.				    \eol
89  Landstreet J.D. et al., 1979.		    \eol
90  Hockey M.S., 1971.				    \eol
91  Heuvel E.P.J. van den, 1971.		    \eol
92  Pilachowski C.A. et al., 1974.                  \eol
93  Wood H.J. \& Campusano L.B., 1975.		    \eol
94  Bonsack W.K., 1977. 			    \eol
95  Hensberge G., 1974. 			    \eol
96  Rudy R.J. \& Kemp C.J., 1978.		    \eol
97  Kuvshinov V.M., 1972.			    \eol
98  Kuvshinov V.M. et al., 1976.		    \eol
99  Babcock H.W. \& Burd S., 1952.		    \eol
100  Evans J.C. \& Elste G., 1971.		    \eol
101  Weiss W.W. \& Wood H.J., 1975.		    \eol
102  Weiss W.W. et al., 1978.			    \eol
103  Hensberge H., De Loore C., 1974.		    \eol
104  Oetken L. et al., 1970.			    \eol
105  Scholz G., 1975.				    \eol
106  Scholz G., 1971a.				    \eol
107  Kemp J.C. \& Wolstencroft R.D., 1973c.	    \eol
108  Hildebrandt G. et al., 1973.		    \eol
109  Oetken L. \& Orwert R., 1973.		    \eol
110  Glagolevskij Yu.V. et al., 1977.		    \eol
111  Gollnow H., 1964.				    \eol
112  Glagolevskij Yu.V. et al., 1981a.		    \eol
113  Glagolevskij Yu.V. et al., 1981b.		    \eol
114  Chunakova N.M. et al., 1981b.                  \eol
115  Aslanov I.A. \& Salomatina N.A., 1981.	    \eol
116  Gerth E., 1981.				    \eol
117  Skulskij M.Yu., 1981.			    \eol
118  Rustamov Yu.S. \& Khotnyanskij A.N., 1980.     \eol
119  Adam M.G., 1965.				    \eol
120  Gollnow H., 1971.				    \eol
121  Maitzen H.M. et al., 1980. 		    \eol
122  Landstreet J.D., 1980.			    \eol
123  Glagolevskij Yu.V. et al., 1979.		    \eol
124  Bonsack W.K. \& Simon T., 1982.		    \eol
125  Preston G.W. \& Sturch C., 1967.		    \eol
126  Wolstencroft R.D. et al., 1981.		    \eol
127  Kemp J.C. \& Wolstencroft R.D., 1974.	    \eol
128  Severny A.B., 1970.			    \eol
129  Landstreet J.D. \& Borra E.F., 1975.	    \eol
130  Aslanov I.A. \& Rustamov Y.S., 1975.	    \eol
131  Severny A.B. et al., 1974. 		    \eol
132  Brown D.N. et al., 1981.			    \eol
133  Wood H.J. \& Albrecht R., 1981.		    \eol
134  Babcock H.W., 1967.			    \eol
135  Bohlender D.A. et al., 1987.		    \eol
136  Scholz G. \&  Gerth E., 1980.		    \eol
137  Rudiger G. \& Scholz G., 1988.		    \eol
138  Fahlman G.G. et al., 1974. 		    \eol
139  Gerth E., 1978.				    \eol
140  Glagolevskij Yu.V. et al., 1989.		    \eol
141  Bychkov V.D. et al., 1989. 		    \eol
142  Glagolevskij Yu.V. et al., 1982a.              \eol
143  Merrill P.W., 1959.			    \eol
144  Saar S.H. \& Linsky J.L., 1985a.               \eol
145  Saar S.H. et al., 1986.			    \eol
146  Zverko J. et al., 1989.			    \eol
147  Mikulasek Z. et al., 1984. 		    \eol
148  Glagolevskij Yu.V. et al., 1984c.              \eol
149  Ryabchikova T.A. \& Piskunov N.E., 1984.	    \eol
150  Glagolevskij Yu.V., 1985.			    \eol
151  St\c epie\'n K., 1984.                         \eol
152  Scholz G., 1984.				    \eol
153  Romanov Yu.S. et al., 1988. 		    \eol
154  Glagolevskij Yu.V. et al., 1984a.		    \eol
155  Glagolevskij Yu.V. et al., 1984b.		    \eol
156  Skulskij M.Yu., 1984.			    \eol
157  Romanov Yu.S. et al., 1984.		    \eol
158  Glagolevskij Yu.V. et al., 1988.		    \eol
159  Ryabchikova T.A. et al., 1988.		    \eol
160  Skulskij M.Yu., 1988a.                         \eol
161  Plachinda S.I., 1988.			    \eol
162  Mikulasek Z., 1988.			    \eol
163  Gerth E., 1988.				    \eol
164  Iliev I.Kh. et al., 1988.			    \eol
165  Hubrig S., 1988.				    \eol
166  Willson R.F. et al., 1988. 		    \eol
167  Saar S.H., 1988.				    \eol
168  Thompson I.B. et al., 1987.		    \eol
169  Landstreet J.D., 1988.			    \eol
170  Mathys G., 1987.				    \eol
171  Bohlender D.A., 1989.			    \eol
172  Landstreet J.D., 1990.			    \eol
173  El'kin V.G. et al., 1987.                      \eol
174  Thompson I.B. \& Landstreet J.D., 1985.	    \eol
175  Renson P., 1984.				    \eol
176  Wolff S.C. \& Morrison N.D., 1974. 	    \eol
177  Skulskij M.Yu., 1988b.                         \eol
178  Gerth E., 1990.				    \eol
179  Ziznovsky J. \& Romanyuk I.I., 1990.	    \eol
180  Skulskij M.Yu., 1985.			    \eol
181  Shore S.N. et al., 1990.			    \eol
182  Bohlender D.A. \& Landstreet J.D., 1990a.	    \eol
183  Bohlender D.A. \& Landstreet J.D., 1990b.	    \eol
184  Mathys G., 1991.				    \eol
185  Iliev I.Kh. \& Barzova I.S., 1990. 	    \eol
186  Thompson I.B., 1983.			    \eol
187  Albrecht R. et al., 1977.			    \eol
188  Landstreet J.D. et al., 1989.		    \eol
189  Plachinda S.I., 1990.			    \eol
190  Weiss W.W. et al., 1990.			    \eol
191  Skulskij M.Yu., 1990.			    \eol
192  El'kin V.G., 1990.                             \eol
193  Iliev I.Kh. et al., 1990.			    \eol
194  Gerth E. et al., 1991a.                        \eol
195  Bychkov V.D. et al., 1991. 		    \eol
196  Lebedev V.S., 1990.			    \eol
197  El'kin V.G. et al., 1991.                      \eol
198  Gerth E. et al., 1991b.                        \eol
199  Harmanec P. et al., 1991.			    \eol
200  Barker P.K. et al., 1985.			    \eol
201  Borra E.F., 1994.				    \eol
202  Burnashev V.I. \& Skulskij M.Yu., 1991.	    \eol
203  Glagolevskij Yu.V. et al., 1986.		    \eol
204  Landstreet J.D., 2000.			    \eol
205  Tarasov E.A. et al., 1992. 		    \eol
206  Shtol' V.G., 1991.                             \eol
207  Vetesnik M., 1983. 			    \eol
208  Scholz G., 1983.				    \eol
209  Scholz G., 1978.				    \eol
210  Bopp B.W. et al., 1989.			    \eol
211  Romanyuk I.I., 1991.			    \eol
212  Lestrade J.-F. et al., 1985.		    \eol
213  Giampapa M.S. \& Worden S.P., 1982.	    \eol
214  North P. et al., 1992a.                        \eol
215  Shtol' V.G. et al., 1992a.                     \eol
216  Shtol' V.G. et al., 1992b.                     \eol
217  Romanyuk I.I., 1986.			    \eol
218  Mathys G. \& Lanz T., 1992.		    \eol
219  Gerth E. et al., 1992a.			    \eol
220  Bychkov V.D. et al., 1992. 		    \eol
221  North P. et al., 1992b.                        \eol
222  Kopylova F.G. \& Romanyuk I.I., 1992.	    \eol
223  El'kin V.G., 1992a.                            \eol
224  El'kin V.G., 1992b.                            \eol
225  Iliev I.Kh. et al., 1992.			    \eol
226  Hubrig S., 1992.				    \eol
227  Skulskij M.Yu. et al., 1992.		    \eol
228  Gerth E. et al., 1992b.			    \eol
229  Skulskij M.Yu. et al., 1992.		    \eol
230  Bohlender D.A. et al., 1993.		    \eol
231  Mathys G. \& Stenflo J.O., 1988.		    \eol
232  Plachinda S.I. \& Polosukhina N., 1994.	    \eol
233  El'kin V.G., 1994.                             \eol
234  Gerth E. \& Bychkov V.D., 1994.		    \eol
235  Udovichenko S.N. et al., 1994.		    \eol
236  Romanyuk I.I., 1994.			    \eol
237  Plachinda S.I. et al., 1993.		    \eol
238  Johnstone R.M. \& Penston M.V., 1984.	    \eol
239  Johnstone R.M. \& Penston M.V., 1987.	    \eol
240  Mathys G. \& Lanz T., 1990.		    \eol
241  Glagolevskij Yu.V. et al., 1985b.              \eol
242  Romanov Yu.S. et al., 1994. 		    \eol
243  Glagolevskij Yu.V. et al., 1995.		    \eol
245  Saar S.H. et al., 1985.			    \eol
246  Saar S.H. \& Linsky J.L., 1985b.               \eol
247  Mathys G. \& Lanz T., 1994.		    \eol
248  Wade G.A. et al., 1996.			    \eol
249  Bohlender D.A., 1994.			    \eol
250  Hubrig S. \& Mathys G., 1994.		    \eol
251  Mathys G., 1994a.                              \eol
252  Brown D.N. et al., 1985.			    \eol
253  Wade G.A. et al., 1996.			    \eol
254  Mathys G. et al., 1997.			    \eol
255  Wade G.A. et al., 1996.			    \eol
256  Mathys G. \& Hubrig S., 1997.		    \eol
257  Romanyuk I.I. et al., 1997.		    \eol
258  Alekseev I.Y. \& Gershberg R.E., 1997.	    \eol
259  El'kin V.G. \& Wade G.A., 1997.                \eol
260  Bychkov V.D. et al., 1997. 		    \eol
261  Glagolevskij Yu.V. \& Chuntonov G.A., 1997.    \eol
262  El'kin V.G., 1996.                             \eol
263  Bychkov V.D. et al., 1997a.		    \eol
264  Bychkov V.D. et al., 1997b.		    \eol
265  Bychkov V.D. et al., 1997c.		    \eol
266  Udovichenko S.N. et al., 1997.		    \eol
267  Bychkov V.D. et al., 1997d.		    \eol
268  Bychkov V.D. \& Shtol' V.G., 1997.             \eol
269  Chuntonov G.A., 1997.			    \eol
270  Bychkov V.D. et al., 1997e.		    \eol
271  El'kin V.G. et al., 1997.                      \eol
272  Babel J. \& North P., 1997.		    \eol
274  Wade G.A. et al., 1999.			    \eol
275  El'kin V.G., 1998.                             \eol
276  Romanyuk I.I. \& Kudryavtsev D.O., 1998.	    \eol
277  Babel J. et al., 1995.			    \eol
278  Maitzen H.M. \& Albrecht R., 1975. 		 \eol
279  Steinitz R. \& Pyper D.M., 1970.			 \eol
280  Freedman R.S., 1978.				 \eol
281  Wade G.A. et al., 1997.				 \eol
282  Steinitz R. \& Pyper D.M., 1971.			 \eol
283  Romanyuk I.I. et al., 1998.			 \eol
284  Mathys G., 1993.					 \eol
285  Mathys G., 1994.                                    \eol
286  Glagolevskij Yu.V. \& Chuntonov G.A., 1998. 	 \eol
287  Romanyuk I.I., 1980.				 \eol
288  Romanyuk I.I., 1984.				 \eol
289  El'kin V.G., 1999.                                  \eol
290  Gerth E., 1994.					 \eol
291  Wade G.A. et al., 1997.				 \eol
292  Glagolevskij Yu.V. et al., 1982b.                   \eol
293  Glagolevskij Yu.V. et al., 1984d.                   \eol
294  Glagolevskij Yu.V. et al., 1981.			 \eol
295  Glagolevskij Yu.V. et al., 1978.			 \eol
296  Bedford D.K. et al., 1995. 			 \eol
297  El'kin V.G., 1995.                                  \eol
298  Marcy G.W., 1984.					 \eol
299  Borra E.F. et al., 1984.				 \eol
300  Dudorov A.E., 1995.				 \eol
301  Mathys G., 1995a.					 \eol
302  Mathys G., 1995b.					 \eol
303  Basri G. et al., 1992.				 \eol
304  Johnstone R.M. \& Penston M.V., 1986.		 \eol
306  Takada-Hidai M. \& Jugaku J., 1993.		 \eol
307  Bedford D.K. et al., 1994. 			 \eol
308  Guenther E.W. et al., 1999.			 \eol
309  Hill G.M. \& Blace C.C., 1996.			 \eol
310  Wade G.A. et al., 2000a.                            \eol
311  Plachinda S.I. \& Tarasova T.N., 2000.		 \eol
312  Wade G.A. et al., 2000b.                            \eol
313  Marcy G.W. \& Bruning D.H., 1984.			 \eol
314  Giampapa M.S. \& Golub L., 1983.			 \eol
315  Basri G. \& Marcy G., 1988.			 \eol
316  Valenti J.A. et al., 1995. 			 \eol
317  Bagnulo S. \& Landolfi M., 1999.			 \eol
318  Gerth E. et al., 1999.				 \eol
319  El'kin V.G. et al., 2001.                           \eol
320  Wade G.A. et al., 2000.				 \eol
321  Tarasova T.N., 2002.				 \eol
322  Hubrig S. et al., 1994.				 \eol
323  Mathys G. \& Hubrig S., 1995.			 \eol
324  Leone F. \& Catanzaro G., 2001.			 \eol
325  Verdugo E. et al., 2002.				 \eol
326  El'kin V.G., 2000.                                  \eol
327  Bychkov V.D., Bychkova L.V., 2002. 		 \eol
328  Udovichenko S., Keir L., 1995.                      \eol
329  Elkin V.G. et al., 2002.                            \eol
330  Chuntonov G.A., 2001.                               \eol
331  Glagolevskij Yu.V. \& Chuntonov G.A., 2001.         \eol
332  Bohlender D.A. \& Landstreet J.D., 1990.            \eol
333  Shorlin S.L.S. et al., 2002.                        \eol
334  Monin D.N. et al., 2002.                            \eol
335  Plachinda S.I., 2000.                               \eol
336  Wade G.A. et al., 2002.                             \eol

\end{document}